\documentclass[11pt,a4paper,twoside,groupcitations]{article}
\usepackage[T1]{fontenc}
\usepackage[ansinew]{inputenc}
\usepackage[english]{babel}
\usepackage{amsfonts}
\usepackage{amsmath}
\usepackage{array}
\usepackage{amsthm}
\usepackage{amssymb}
\usepackage{graphicx}
\usepackage{subfigure}
\usepackage{braket}
\usepackage{eucal}
\usepackage{verbatim}
\usepackage[table]{xcolor}
\usepackage{caption}
\usepackage{cite}
\usepackage{textcomp}
\usepackage[shortlabels]{enumitem}
\newcommand{\be}{\begin{eqnarray}}
\newcommand{\ee}{\end{eqnarray}}

\title{\bf A causal Schwarzschild-de~Sitter interior solution\\ by gravitational decoupling}
\author{L.~Gabbanelli$^{a}$\thanks{gabbanelli@icc.ub.edu}
$\,$,
J.~Ovalle$^{bc}$\thanks{jovalle@usb.ve}
$\,$, 
A.~Sotomayor$^{d}$\thanks{adrian.sotomayor@uantof.cl}
$\,$, 
Z.~Stuchlik$^{b}$\thanks{zdenek.stuchlik@fpf.slu.cz}
$\,$,
R.~Casadio$^{ef}$\thanks{casadio@bo.infn.it}
\\
\null
\\
$^a${\em Departament de F\'{\i}sica Qu\`antica i Astrof\'{\i}sica, Institut de Ci\`encies del Cosmos (ICCUB),}
\\
{\em Universitat de Barcelona, Mart\'i i FranquÃ¨s 1, 08028 Barcelona, Spain}
\\
$^b${\em Institute of Physics and Research Centre of Theoretical Physics and Astrophysics,}
\\
{\em Faculty of Philosophy and Science, Silesian University in Opava}
\\
{\em CZ-746 01 Opava, Czech Republic}
\\
$^c${\em Departamento de F\'{\i}sica, Universidad Sim\'on Bol\'ivar,}
\\
{\em AP 89000, Caracas 1080A, Venezuela}
\\
$^d${\em Departamento de Matem\'aticas, Universidad de Antofagasta}
\\
{\em  Antofagasta, Chile}
\\
$^e${\em Dipartimento di Fisica e Astronomia, Alma Mater Universit\`a di Bologna}
\\
{\em via Irnerio~46, 40126 Bologna, Italy}
\\
$^f${\em Istituto Nazionale di Fisica Nucleare, Sezione di Bologna, I.S.~FLAG}
\\
{\em  viale Berti~Pichat~6/2, 40127 Bologna, Italy}
}
\begin{document}
\maketitle
\begin{abstract}
We employ the minimal geometric deformation approach to gravitational decoupling (MGD-decoupling)
in order to build an exact anisotropic version of the Schwarzschild interior solution in a space-time with cosmological
constant.
Contrary to the well-known Schwarzschild interior, the matter density in the new solution is not uniform
and possesses subluminal sound speed.
It therefore satisfies all standard physical requirements for a candidate astrophysical object.
\end{abstract}
%
%
%
%
%
%
%
\section{Introduction}
\label{S:intro}
\setcounter{equation}{0}
The Schwarzschild interior metric is one of the best known solutions of Einstein's field equations~\cite{Schwarzschild}.
This exact solution represents an isotropic self-gravitating object of uniform density (incompressible fluid),
and has been widely studied, generally without considering the cosmological constant.
As far as we know, it is one of the few analytic solutions for a bounded distribution which fits smoothly with the Schwarzschild
exterior metric~\cite{weinberg}. 
However, it cannot be used to represent a stellar model as its speed of sound is not subliminal (see Ref.~\cite{som}
where a model of two fluids is used to circumvent the problem of causality).
Despite of the above, some studies on the possible impact of the vacuum energy on perfect fluids have been carried
out which made use of this solution for both positive and negative values of the cosmological constant.
Such analyses can help to elucidate some properties of the Schwarzschild-(anti-)de Sitter space-time in presence of
matter~\cite{zdenek1,zdenek2} (see also~\cite{zdenek3,bohmer,zdenek4}).
Moreover, some alternatives to black holes, like the gravastars~\cite{mazur1,mazur2}, are mainly generated from
this solution~\cite{camilo1,camilo2}, and an exact time-dependent version was recently reported in Ref.~\cite{gondolo}.
Therefore, it is not only natural, but also useful, to construct a possible extension of this solution for more realistic stellar
scenarios, such as that represented by non-uniform and anisotropic self-gravitating objects.
Above all, it would be very important to develop versions that do not suffer from the causal problem.
However, given the complexity of Einstein's field equations, we know that extending a known solution to more complex
scenarios is an arduous and difficult task~\cite{Stephani}, even more so if we wish to keep it physically acceptable.
Fortunately, the so-called method of gravitational decoupling by Minimal Geometric Deformation
(MGD-decoupling, henceforth)~\cite{MGD-decoupling,MGDe-decoupling}, which has been widely used
recently~\cite{mgdaniso,Luciano,MGDDBH,sharif1,Inverse,sharif2,ernestopedro,tello,camilo,angel2,sharif3,
ernestoangelpedro,mauryatello,ernestocqg1,EKG,sharif4},
has proved to be a powerful method to extend known solutions into more complex scenarios.
\par
The original version of the MGD approach was developed in Refs.~\cite{jo1,jo2} in the context of the brane-world
scenario~\cite{lisa1,lisa2}, and it was eventually extended to study black hole solutions in
Refs.~\cite{MGDextended1,MGDextended2} (for some earlier works on the MGD, see for instance
Refs.~\cite{jo6,jo8,jo9,jo10},
and Refs.~\cite{jo11,jo12,roldaoGL,rrplb,rr-glueball,rr-acustic,fr,MGDBH,emgd,ffroldao} for some recent applications).
On the other hand, the MGD-decoupling has three main characteristics that make it particularly useful in the search
for new solutions of Einstein's field equations, namely:
\begin{enumerate}[I.]
\item
one can extend simple solutions of the Einstein equations into more complex domains. 
In fact, we can start from a source with energy-momentum tensor $\hat T_{\mu\nu}$ for which the metric is known
and add the energy-momentum tensor of a second source,
\begin{equation}
\label{coupling0}
\hat T_{\mu\nu}
\rightarrow
T_{\mu\nu}
=
\hat T_{\mu\nu}
+T^{(1)}_{\mu\nu}
\ .
\end{equation}
We can then repeat the process with more sources $T^{(i)}_{\mu\nu}$ to extend the solution of the Einstein equations
associated with the gravitational source $\hat T_{\mu\nu}$ into the domain of more intricate forms of gravitational sources
$T_{\mu\nu}$;
\item
one can also reverse the previous procedure in order to find a solution to Einstein's equations with a complex
energy-momentum tensor ${T}_{\mu\nu}$ by splitting it into simpler components, 
\begin{equation}
\label{split}
{T}_{\mu\nu}
\rightarrow
\hat T_{\mu\nu}+T^{(i)}_{\mu\nu}
\ ,
\end{equation} 
and solve Einstein's equations for each one of these components.
Hence, we will have as many solutions as the components in the original energy-momentum tensor ${T}_{\mu\nu}$.
Finally, by a simple combination of all these solutions, we will obtain the solution to the Einstein equations associated
with the original energy-momentum tensor ${T}_{\mu\nu}$.
We emphasise that the MGD-decoupling works as long as the sources do not exchange energy-momentum directly
among them,
to wit
\begin{equation}
\nabla_{\mu}\hat T^{\mu\nu}
=
\nabla_{\mu} T^{(1)\mu\nu}
=
\ldots
=
\nabla_{\mu} T^{(n)\mu\nu}
=
0
\ ,
\label{nablas}
\end{equation}
which further clarifies that their interaction is purely gravitational;
\item
it can be applied to theories beyond general relativity.
For instance, given the modified action~\cite{MGDe-decoupling}
\begin{equation}
\label{ngt}
S_{\rm G}
=
S_{\rm EH}+S_{\rm X}
=
\int\left[\frac{R}{2\,k^2}+{\cal L}_{\rm M}+{\cal L}_{\rm X}\right]\sqrt{-g}\,d^4\,x
\ ,
\end{equation}
where ${\cal L}_{\rm M}$ contains all matter fields in the theory and ${\cal L}_{\rm X}$ is the Lagrangian density
of a new gravitational sector with an associated (effective) energy-momentum tensor
\begin{equation}
\label{ngt2}
\theta_{\mu\nu}
=
\frac{2}{\sqrt{-g}}\frac{\delta(\sqrt{-g}\,{\cal L}_{\rm X})}{\delta g^{\mu\nu}}
=
2\,\frac{\delta{\cal L}_{\rm X}}{\delta g^{\mu\nu}}-g_{\mu\nu}\,{\cal L}_{\rm X}
\ ,
\end{equation}
the method in I.~allows one to extend all the known solutions of the Einstein-Hilbert action $S_{\rm EH}$
into the domain of modified gravity represented by $S_{\rm G}$.
This represents a straightforward way to study the consequences of extended gravity on general relativity. 
\end{enumerate}
In this paper we will apply the procedure I.~to the Schwarzschild interior solution in order to build a new interior
configuration with non-uniform matter density and anisotropic pressure.
\par
The paper is organised as follows:
in Section~\ref{s2}, we start from the Einstein equations with cosmological constant for a spherically symmetric
stellar distribution and we show how to decoupling two spherically symmetric and static gravitational sources
$\{T_{\mu\nu},\,\theta_{\mu\nu}\}$.
After providing details on the matching conditions at the star surface under the MGD-decoupling,
in Section~\ref{s5}, we implement the MGD-decoupling following the scheme I.~to generate the extended
version of the Schwarzschild solution; finally, we summarise our conclusions in Section~\ref{con}.
\section{Spherically symmetric stellar distribution}
\label{s2}
\setcounter{equation}{0}
Let us start from the standard Einstein field equations~\footnote{We use the metric signature $(+---)$
and the constant $k^2=8\,\pi\,G_{\rm N}$.}
\begin{equation}
\label{EinEq}
R_{\mu\nu}-\frac{1}{2}\,R\,g_{\mu\nu}
+\Lambda \,g_{\mu\nu}
=
k^2\,T_{\mu\nu}
\ ,
\end{equation}
where $\Lambda$ is a positive cosmological constant.
The energy-momentum tensor $T_{\mu\nu}$ in Eq.~\eqref{EinEq} is given by
\begin{equation}
\label{emt}
T_{\mu\nu}
=
\hat{T}_{\mu\nu}+\theta_{\mu\nu}
\ ,
\end{equation}
where $\hat{T}_{\mu\nu}$ represents the energy-momentum tensor of a perfect fluid, and $\theta_{\mu\nu}$
adds anisotropic effects on ${T}_{\mu\nu}$.
Since the Einstein tensor is divergence free, the total energy momentum tensor ${T}_{\mu\nu}$ must satisfy
the conservation equation
\begin{equation}
\nabla_\nu\,T^{{\mu\nu}}
=
0
\ .
\label{dT0}
\end{equation}
In Schwarzschild-like coordinates, the spherically symmetric metric reads 
\begin{equation}
ds^{2}
=
e^{\nu (r)}\,dt^{2}-e^{\lambda (r)}\,dr^{2}
-r^{2}\left( d\theta^{2}+\sin ^{2}\theta \,d\phi ^{2}\right)
\ ,
\label{metric}
\end{equation}
where $\nu =\nu (r)$ and $\lambda =\lambda (r)$ are functions of the areal radius $r$ only, ranging from $r=0$
(the star's centre) to some $r=R>0$ (the star's surface).
The cosmological constant can be thought to contribute the stress-energy tensor being responsible for
the expansion of the universe, with a non-zero vacuum energy density and negative pressure satisfying
$\rho_{vac}=-p_{vac}=\Lambda/k^2$.
Explicitly, the field equations read
\begin{eqnarray}
\label{ec1}
&&
k^2
\left( \rho
+\theta_0^{\,0}
\right)+\Lambda
=
\strut\displaystyle\frac 1{r^2}
-e^{-\lambda }\left( \frac1{r^2}-\frac{\lambda'}r\right)\ ,
\\
&&
\label{ec2}
k^2
\strut\displaystyle
\left(-p+\theta_1^{\,1}\right)+\Lambda
=
\frac 1{r^2}-e^{-\lambda }\left( \frac 1{r^2}+\frac{\nu'}r\right)\ ,
\\
&&
\label{ec3}
k^2
\strut\displaystyle
\left(-p+\theta_2^{\,2}\right)+\Lambda
=
\frac 14e^{-\lambda }\left[ -2\,\nu''-\nu'^2+\lambda'\,\nu'
-2\,\frac{\nu'-\lambda'}r\right]
\ ,
\end{eqnarray}
while the conservation equation, which is a linear combination of Eqs.~(\ref{ec1})-(\ref{ec3}), yields
\begin{equation}
\label{con1}
-p'-\strut\displaystyle\frac{\nu'}{2}\left(\rho+p\right)
+(\theta_1^{\,\,1})'-\strut\displaystyle\frac{\nu'}{2}\left(\theta_0^{\,\,0}
-\theta_1^{\,\,1}\right)-\frac{2}{r}\left(\theta_2^{\,\,2}-\theta_1^{\,\,1}\right)
=
0
\ ,
\end{equation}
where $f'\equiv \partial_r f$.
\par
By simple inspection of Eqs.~(\ref{ec1})-(\ref{ec3}), we can identify an effective density 
\begin{equation}
\tilde{\rho}
=
\rho
+\theta_0^{\,0}
\ ,
\label{efecden}
\end{equation}
an effective isotropic pressure
\begin{equation}
\tilde{p}_{r}
=
p-\theta_1^{\,1}
\ ,
\label{efecprera}
\end{equation}
and an effective tangential pressure
\begin{equation}
\tilde{p}_{t}
=
p-\theta_2^{\,2}
\ .
\label{efecpretan}
\end{equation}
This clearly illustrate that the source $\theta_{\mu\nu}$ generates an anisotropy 
\begin{equation}
\label{anisotropy}
\Pi
\equiv
\tilde{p}_{t}-\tilde{p}_{r}
=\theta_1^{\,1}-\theta_2^{\,2}
\end{equation}
inside the stellar distribution.  
\par
Eqs.~(\ref{ec1})-(\ref{ec3}) contain five unknown functions, namely, 
three physical variables: the density $\tilde{\rho}(r)$, the radial pressure $\tilde{p}_r(r)$
and the tangential pressure $\tilde{p}_t(r)$; and two geometric functions:
the temporal metric function $\nu(r)$ and the radial metric function $\lambda(r)$.
Therefore these equations form an indefinite system~\cite{Luis,tiberiu} which requires
additional information to produce any specific solution.
\subsection{Gravitational decoupling by MGD}
\label{s3}
In order to solve the Einstein equations~(\ref{ec1})-(\ref{con1}) we implement the MGD-decoupling. 
In this approach, one starts from a solution for the isotropic fluid and the field equations with the anisotropic source
$\theta_{\mu\nu}$ will take the form of effective ``quasi-Einstein'' equations [see Eqs.~(\ref{ec1d})-(\ref{ec3d}) below].
\par
A solution to Eqs.~(\ref{ec1})-(\ref{con1}) with $\theta_{\mu\nu}=0$ will be given by a General Relativity perfect fluid solution
with cosmological constant $\Lambda$ and be characterised by the four functions $\{\xi,\mu,\rho,p\}$ such that 
the metric reads
\begin{equation}
ds^{2}
=
e^{\xi (r)}\,dt^{2}
-
e^{\mu(r)}\,dr^{2}
-
r^{2}\left( d\theta^{2}+\sin ^{2}\theta \,d\phi ^{2}\right)
\ ,
\label{pfmetric}
\end{equation}
where 
\begin{equation}
\label{standardGR}
e^{-\mu(r)}
\equiv
1-\frac{k^2}{r}\int_0^r x^2\,\rho\, dx
=
1-\frac{2\,m(r)}{r}
\end{equation}
is the standard General Relativity expression containing the Misner-Sharp mass function $m=m(r)$.
Next, we turn on the anisotropic effects by adding the $\theta_{\mu\nu}$.
These effects can be encoded in the geometric deformation undergone by the
perfect fluid geometry in Eq.~(\ref{pfmetric}), namely
\begin{eqnarray}
\label{gd1}
\xi
&\mapsto &
\nu
=
\xi+\alpha\,g
\ ,
\\
\label{gd2}
e^{-\mu}
&\mapsto &
e^{-\lambda}
=
e^{-\mu}+\alpha\,f
\ ,
\end{eqnarray}
where $g$ and $f$ are the deformations undergone by the temporal and radial
metric component of the perfect fluid geometry $\{\xi,\mu\}$, respectively.
Among all possible deformations~(\ref{gd1}) and (\ref{gd2}), the so-called minimal geometric
deformation is given by $g=0$ and $f=f^*$, where $f^*$ satisfies a suitable condition~\cite{jo1,jo2}
in order to minimise the departure from General Relativity.
Only the radial metric component therefore changes to
\begin{eqnarray}
\label{expectg}
e^{-\mu(r)}\mapsto\,e^{-\lambda(r)}
=
e^{-\mu(r)}+\alpha\,f^{*}(r)
\ .
\end{eqnarray}
The system~\eqref{ec1}-\eqref{con1} can be decoupled by plugging the deformation~(\ref{expectg})
into the Einstein equations~(\ref{ec1})-(\ref{ec3}).
In fact, the system splits into two sets of equations:
(i) one having the standard Einstein field equations for a perfect fluid with cosmological constant $\Lambda$,
whose metric is given by Eq.~(\ref{pfmetric}) with $\xi(r)=\nu(r)$,
\begin{eqnarray}
\label{ec1pf}
&&
k^2\rho+\Lambda
=
\strut\displaystyle\frac 1{r^2}
-e^{-\mu }\left( \frac1{r^2}-\frac{\mu'}r\right)\ ,
\\
&&
\label{ec2pf}
k^2
\left(-p\right)+\Lambda
=
\frac 1{r^2}-e^{-\mu }\left( \frac 1{r^2}+\frac{\nu'}r\right)\ ,
\\
&&
\label{ec3pf}
k^2
\strut\displaystyle
\left(-p\right)
+\Lambda
=
\frac 14e^{-\mu }\left[ -2\,\nu''-\nu'^2+\mu'\,\nu'
-2\,\frac{\nu'-\mu'}r\right]
\ ,
\end{eqnarray}
along with the conservation equation~(\ref{dT0}) with $\theta_{\mu\nu} = 0$, 
namely $\nabla_\nu\,\hat{T}^{{\mu\nu}}=0$, yielding
\begin{equation}
\label{conpf}
p'+\strut\displaystyle\frac{\xi'}{2}\,(\rho+p)
=
0
\ ,
\end{equation}
which is a linear combination of Eqs~(\ref{ec1pf})-(\ref{ec3pf});
and (ii) one for the source $\theta_{\mu\nu}$, which reads
\begin{eqnarray}
\label{ec1d}
&&
k^2\,\theta_0^{\,0}
=
-\strut\displaystyle\frac{\alpha\,f^{*}}{r^2}
-\frac{\alpha\,f^{*'}}{r}\ ,
\\
&&
\label{ec2d}
k^2
\strut\displaystyle
\,\theta_1^{\,1}
=- \alpha\,f^{*}\left(\frac{1}{r^2}+\frac{\nu'}{r}\right)\ ,
\\
&&
\label{ec3d}
k^2
\strut\displaystyle\,\theta_2^{\,2}
=
-\frac{\alpha\,f^{*}}{4}\left(2\nu''+\nu'^2+\frac{2\,\nu'}{r}\right)-\frac{\alpha\,f^{*'}}{4}\left(\nu'+\frac{2}{r}\right)
\ .
\end{eqnarray}
The conservation equation  $\nabla_\nu\,\theta^{\mu\nu}=0$ explicitly reads
\begin{equation}
\label{con1d}
(\theta_1^{\,\,1})'
-\strut\displaystyle\frac{\nu'}{2}(\theta_0^{\,\,0}
-\theta_1^{\,\,1})-\frac{2}{r}(\theta_2^{\,\,2}-\theta_1^{\,\,1})
=
0
\ ,
\end{equation}
which is a linear combination of Eqs.~(\ref{ec1d})-(\ref{ec3d}). 
We recall that, under these conditions, there is no exchange of energy-momentum between the perfect fluid
and the source $\theta_{\mu\nu}$ and therefore  their interaction is purely gravitational. 
\subsection{Matching conditions at the surface}
\label{s4}
The interior ($0\le r\le R$) of the self-gravitating system of radius ($r=R$) is described by the MGD metric
\begin{equation}
ds^{2}
=
e^{\nu^{-}(r)}\,dt^{2}
-\left[1-\frac{2\,\tilde{m}(r)}{r}\right]^{-1}dr^2
-r^{2}\left(d\theta ^{2}+\sin {}^{2}\theta d\phi ^{2}\right)
\ ,
\label{mgdmetric}
\end{equation}
where the interior mass function is given by
\begin{equation}
\label{effecmass}
\tilde{m}(r)
=
m(r)-\frac{r}{2}\,\alpha\,f^{*}(r)
\ , 
\end{equation} 
with the Misner-Sharp mass $m$ given in Eq.~(\ref{standardGR}) and $f^{*}$ the geometric deformation in Eq.~(\ref{expectg}).
On the other hand, the exterior ($r>R$) space-time will be described by the Schwarzschild-de~Sitter metric  
\begin{equation}
\label{MetricSdS}
ds^2
=
\left(1-\frac{2\,{\cal M}}{r}-\frac{\Lambda}{3}\,r^2\right)\,dt^2
-\frac{dr^2}{\left(1-\frac{2\,{\cal M}}{r}-\frac{\Lambda}{3}\,r^2\right)}
-r^{2}\left( d\theta^{2}+\sin ^{2}\theta \,d\phi ^{2}\right)\ . 
\end{equation}
\par
The metrics in Eqs.~\eqref{mgdmetric} and~\eqref{MetricSdS} must satisfy the Israel-Darmois matching
conditions~\cite{israel} at the star surface $\Sigma$ defined by $r=R$.
In particular, the continuity of the metric across $r=R$ implies
\begin{equation}
e^{\nu ^{-}(R)}
=
1-\frac{2\,{\cal M}}{R}-\frac{\Lambda}{3}\,R^2
\ ,
\label{ffgeneric1}
\end{equation}
and
\begin{equation}
1-\frac{2\,M}{R}+\alpha\,f^{*}_{R}
=
1-\frac{2\,{\cal M}}{R}
\ ,
\label{ffgeneric2}
\end{equation}
where $M=m(R)$ and $f^{*}_{R}=f^*(R)$ is the minimal geometric deformation evaluated at the star surface.
\par
Likewise, the extrinsic curvature (or second fundamental form) of spheres
\begin{equation}
K_{\mu\nu}=\nabla_\mu\,r_\nu
\ ,
\end{equation}
where $r_\mu$ is the unit radial vector normal to a surface of constant $r$, must be continuous 
across the sphere $\Sigma$, which can be written in terms of the Einstein tensor
as~\footnote{For more details see for instance Ref.~\cite{santos}.}
\begin{equation}
\left[G_{\mu \nu }\,r^{\nu }\right]_{\Sigma }
\equiv
\lim_{r\to R^+}\left(G_{\mu \nu }\,r^{\nu }\right)
-
\lim_{r\to R^-}\left(G_{\mu \nu }\,r^{\nu }\right)
=
0
\ ,
\label{matching1}
\end{equation}
Using Eq.~(\ref{matching1}) and the general Einstein equations~(\ref{EinEq}),
we then find 
\begin{equation}
\left[(k^2\,T_{\mu \nu }+\Lambda\,g_{\mu\nu})\,r^{\nu }\right]_{\Sigma}
=
0
\ ,
\label{matching2x}
\end{equation}
which leads to 
\begin{equation}
\left[k^2\,\left(p-\theta_1^{\,\,1}\right)-\Lambda\right]_{\Sigma }
=
0
\ .
\label{matching3}
\end{equation}
This matching condition takes the final form 
\begin{equation}
p_{R}
-(\theta_1^{\ 1})^{-}_{R}
=
0
\ ,
\label{matchingf}
\end{equation}
where $p_{R}\equiv p(R)$ and $(\theta_1^{\ 1})^{-}_{R}\equiv \theta_1^{\ 1}({r\to R^-})$.
The condition~(\ref{matchingf}) holds in general for the second fundamental form associated
with the Einstein equations~(\ref{EinEq}) and the energy-momentum~\eqref{emt}.
After decoupling the source $\theta_{\mu\nu}$ and by using Eq.~(\ref{ec2d}) for the interior geometry,
Eq.~(\ref{matchingf}) can be written as
\begin{equation}
\tilde{p}_R\equiv\,p_{R}+\alpha\,\frac{f_{R}^{\ast }}{k^2}
\left(\frac{1}{R^{2}}+\frac{\nu _{R}^{\prime }}{R}\right)=0
\ ,
\label{sfgeneric}
\end{equation}
where $\nu _{R}^{\prime }\equiv \partial _{r}\nu^{-}|_{r=R}$. Eqs.~(\ref{ffgeneric1}), (\ref{ffgeneric2})
and~(\ref{sfgeneric}) are the necessary and sufficient conditions for matching the interior
MGD metric~(\ref{mgdmetric}) with the outer Schwarzschild-de~Sitter metric~\eqref{MetricSdS}.
\par
The expression in Eq.~\eqref{sfgeneric} in particular contains critical information about the conditions that
the self-gravitating system must fulfil in order to be consistently coupled with the Schwarzschild-de~Sitter
geometry~\eqref{MetricSdS}.
First of all, the effective radial pressure $\tilde{p}$ at the surface must vanish, which is a very well-known result.
However, if the geometric deformation $f^*(r<R)$ is positive, hence weakening the gravitational field,
[see Eq.~(\ref{effecmass})], the exterior geometry~\eqref{MetricSdS} can only be compatible with a
non-vanishing $\theta_{\mu\nu}$ if the perfect fluid has $p_{R}<0$, which may be interpreted as regular
matter with a solid crust~\cite{jo11}.
If we want to avoid having a solid-crust and keep the standard condition $p_{R}=0$, we must impose
that the anisostropic effects on the radial pressure pressure vanish at $r=R$.
For instance, this is achieved if we assume that $(\theta_1^{\,\,1})^{-}_{R}\sim\,p_{R}$
in Eq.~(\ref{matchingf}), which leads to a vanishing inner deformation $f^*_R=0$
[see further Eq.~\eqref{Constrain}].
\section{Anisotropic Schwarzschild-de~Sitter interior solution}
\setcounter{equation}{0}
\label{s5}
Let us recall that the interior Schwarzschild-de~Sitter solution for the system~(\ref{ec1pf})-(\ref{conpf}) 
represents a stellar system of radius $R$ formed by an incompressible perfect fluid of total mass $M$ 
and is given by~\cite{zdenek3}
\begin{eqnarray}
\label{Schw00}
&&
e^{\nu(r)}
=\left[A-B\,\sqrt{1-\frac{r^2}{C^2}}\right]^2
\\
&&
\label{Schw11}
e^{-\mu(r)}
=
1-\frac{r^2}{C^2}
\\
&&
\label{Schwden}
k^2\,\rho
=
\frac{3}{C^2}-\Lambda
\\
&&
\label{Schwpre}
k^2\,p(r)
=
\frac{A\left(1-\Lambda\,C^2\right)
-B\left(3-\Lambda\,C^2\right)
\sqrt{1-\frac{r^2}{C^2}}}
{C^2\left(B\sqrt{1-\frac{r^2}{C^2}}-A\right)}
\ .
\end{eqnarray} 
The constants $A$, $B$ and $C$ can be expressed in terms of the physical parameters
$M$, $R$ and $\Lambda$ by means of the matching conditions with the outer
Schwarzschild-de~Sitter vacuum~\eqref{MetricSdS}.
In fact, Eqs.~(\ref{ffgeneric1}), (\ref{ffgeneric2}) and~(\ref{sfgeneric}) with $\alpha=0$ yield
${\cal M}=M$ and
\begin{eqnarray}
\label{Aconstant}
&&
A
=
B\,\frac{3-\Lambda\,C^2}{1-\Lambda\,C^2}\,\sqrt{1-\frac{R^2}{C^2}}
\\
\label{Bconstant}
&&
B
=
\frac{3\,M-\Lambda\,R^3}{6\,M+\Lambda\,R^3}
\\
\label{RCconstant}
&&
\frac{R^2}{C^2}
=
\frac{2\,M}{R}+\frac{\Lambda\,R^2}{3}
\ .
\end{eqnarray}
We see that the pressure~\eqref{Schwpre} becomes singular in the centre $r=0$ when 
\begin{equation}
\label{AB}
A=B
\ .
\end{equation}
We can avoid the singularity, while we keep $p(0)>0$, when the compactness satisfies
\begin{equation}
\label{buch}
\frac{M}{R}<\frac{1}{9}\left(2+\sqrt{4-3\,\Lambda\,R^2}\right)
\ ,
\end{equation}
which is the Buchdahl inequality with cosmological constant (for the details, see for instance, Ref.~\cite{cb}).
For simplicity, we just show the final expressions for the case $\Lambda=0$~\footnote{From the phenomenological
point of view, the cosmological constant $\Lambda\sim\,10^{-52}$ appears irrelevantly small for any 
astrophysical systems.},
for which we have
\begin{eqnarray}
\label{Schw000}
&&
e^{\nu(r)}
=
\frac{1}{4}
\left(
\sqrt{1-\frac{2\,M\,r^2}{R^3}}
-3\,\sqrt{1-\frac{2\,M}{R}}
\right)^2
\\
&&
\label{Schw110}
e^{-\mu(r)}
=
1-\frac{2\,M\,r^2}{R^3}
\\
&&
\label{Schwpre0}
k^2\,p(r)
=
\frac{6\,M\left(\sqrt{R^3-2\,M\,r^2}-R\,\sqrt{R-2\,M}\right)}
{R^3\left(3\,R\,\sqrt{R-2\,M}-\sqrt{R^3-2\,M\,r^2}\right)}
\ ,
\end{eqnarray} 
from which we easily see that $p_R=0$, as required.
\par
In order to generate the anisotropic version of the Eqs.~\eqref{Schw00}-\eqref{Schwpre},
we need to determinate the anisotropic source $\theta_{\mu\nu}$,
whose field equations are given by the expressions~\eqref{ec1d}-\eqref{ec3d}.
This system has four unknowns and we therefore need to give some physically motivated prescriptions.
Since we further want to avoid having a solid crust at the surface, we require the radial pressure satisfies
the ``mimic constraint'' defined by 
\begin{equation}
\label{Constrain}
\theta_1^{\,1}(r)
=
\alpha\,p(r)
\ ,
\end{equation}
which, according to Eqs.~\eqref{ec2pf} and~\eqref{ec2d}, yields
\begin{equation}
\label{mimic}
f^*(r)
=
-e^{-\mu(r)}+\frac{1-\Lambda\,r^2}{1+r\,\nu'(r)}
\ .
\end{equation}
 We remark that the constraint~\eqref{Constrain} ensures that the effective radial pressure vanishes at
$r=R$, since $\tilde p(R)=p(R)=0$ from Eq.~\eqref{Schwpre}.
\par
Using the expression~\eqref{mimic} in the MGD deformation~\eqref{expectg}, the radial metric component becomes
\begin{eqnarray}
e^{-\lambda(r)}
&\!\!=\!\!&
\left(1-\alpha\right)
e^{-\mu(r)}
+\alpha\,\frac{1-\Lambda\,r^2}{1+r\,\nu'(r)}
\nonumber
\\
\label{schw11d}
&\!\!=\!\!&
\left(1-\frac{r^2}{C^2}\right)
+\alpha\,\frac{r^2}{C^2}
\left[\frac{A\,(1-\Lambda\,C^2)-B\,(3-\Lambda\,C^2)\,\sqrt{1-\frac{r^2}{C^2}}}
{A\,\sqrt{1-\frac{r^2}{C^2}}-B\left(1-\frac{3\,r^2}{C^2}\right)}\right]
\sqrt{1-\frac{r^2}{C^2}}
\ ,
\end{eqnarray}
where the minimal geometric deformation is given by
\begin{equation}
\label{f}
f^*(r)
=
\frac{r^2}{C^2}
\left[\frac{A\,(1-\Lambda\,C^2)-B\,(3-\Lambda\,C^2)\,\sqrt{1-\frac{r^2}{C^2}}}
{A\sqrt{1-\frac{r^2}{C^2}}-B\,\left(1-\frac{3\,r^2}{C^2}\right)}\right]\sqrt{1-\frac{r^2}{C^2}}
\ .
\end{equation}
\begin{figure}[t]
\center
\includegraphics[scale=0.8]{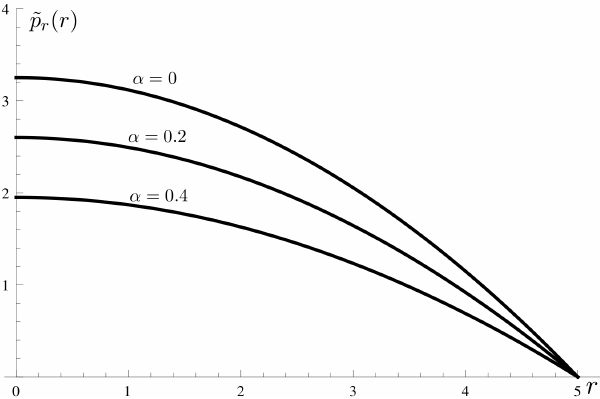}
\includegraphics[scale=0.8]{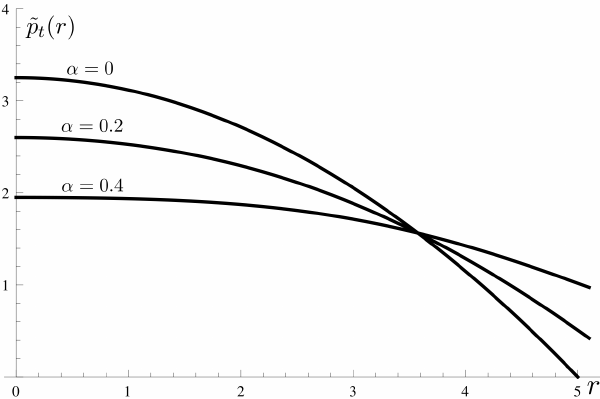}
\\
\caption{Effective radial pressure $\tilde{p}_r(r)$ [$10^{-4}$] and effective tangential pressure
$\tilde{p}_t(r)$ [$10^{-4}$] for a stellar system with compactness $M/R=0.2$ compared
to the standard Schwarzschild case $\alpha=0$ with $p_r=p_t$.
Radial coordinates are in units of $M$ ($r=5=R)$.
The cosmological constant $\Lambda\sim\,10^{-52}$ is too small to be relevant.}
\label{figpr}      
\end{figure}
\par
By using the metric functions $\nu(r)$ and $\lambda(r)$ in~\eqref{Schw00} and~\eqref{schw11d}
in the field equations~(\ref{ec1})-(\ref{ec3}), we find the effective density
\be
\label{Schwd}
k^2\,\tilde{\rho}(r)
=
k^2\,\rho
-\alpha\left(\frac{f^{*}(r)}{r^2}+\frac{f^{*'}(r)}{r}\right)
\ ,
\ee
the effective radial pressure
\be
\label{Schwpr}
k^2\,\tilde{p}_r(r)
=
(1-\alpha)\,k^2\,p(r)
\ ,
\ee
and the effective tangential pressure
\be
\label{Schwpt}
\tilde{p}_t(r)
=
\tilde{p}_r(r)+\Pi(\alpha,r)
\ , 
\ee
where
\be
\Pi(\alpha,r)
&\!\!=\!\!&
\alpha\,
\frac{A\left(1-\frac{r^2}{C^2}\right)^{3/2}
-B \left(1-\frac{2r^2}{C^2}+\frac{2r^4}{C^4}\right)}
{k^2\,r^2\left(1-\frac{r^2}{C^2}\right)^{3/2}
\left(B \sqrt{1-\frac{r^2}{C^2}}-A\right)}
\,f^{*}(r)
\nonumber
\\
&&
+
\alpha\,
\frac{A\, \sqrt{1-\frac{r^2}{C^2}}-B\left(1-\frac{2r^2}{C^2}\right)}
{2\, k^2\, r \left[A\sqrt{1-\frac{r^2}{C^2}}-B\left(1-\frac{r^2}{C^2}\right)\right]}
\,f^{*'}(r)
\ ,
\end{eqnarray}
and is rather cumbersome to display explicitly.
The expressions~(\ref{Schw00}),~(\ref{schw11d}),~(\ref{Schwd})-(\ref{Schwpt})
represent an exact solution for the anisotropic system~(\ref{ec1})-(\ref{ec3}).
\begin{figure}[t]
\center
\includegraphics[scale=0.8]{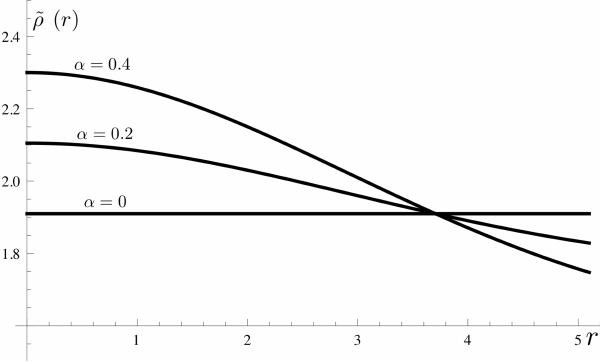}
\includegraphics[scale=0.8]{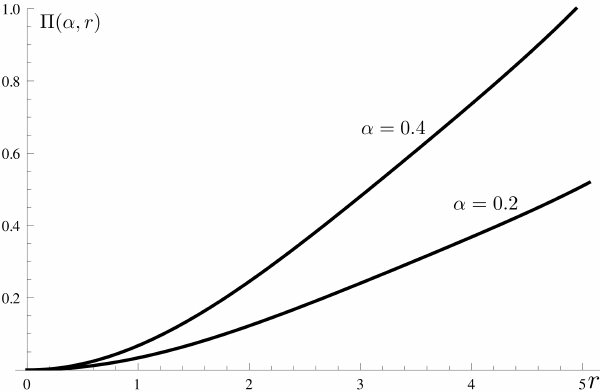}
\\
\caption{Effective density $\tilde{\rho}(r)$ [$10^{-3}$] and anisotropy $\Pi(\alpha,r)$ [$10^{-4}$]
for a stellar system with compactness $M/R=0.2$.
Radial coordinates are in units of $M$ ($r=5=R)$.
The cosmological constant $\Lambda\sim\,10^{-52}$ is too small to be relevant.}
\label{figrho}      
\end{figure}
\par
We can next match the interior metric~(\ref{metric}) with metric functions~(\ref{Schw00})
and (\ref{schw11d}) with the exterior Schwarzschild-de~Sitter vacuum solution~({\ref{MetricSdS}}).
We can see that, for a given total mass $M$ and radius $R$, we have four unknown parameters,
namely $A$, $B$ and $C$ from the interior solution in Eqs.~(\ref{Schw00}) and (\ref{schw11d}), and the mass
${\cal M}$ in Eq.~({\ref{MetricSdS}}).
However, the mass $M$ is related to the constant $C$ and the radius $R$ by the definition~\eqref{standardGR}. 
We therefore have only three unknown constants to be determined by the three conditions~(\ref{ffgeneric1}),
(\ref{ffgeneric2}) and~(\ref{matchingf}) at the star surface.
The continuity of the metric given by Eqs.~(\ref{ffgeneric1}) and (\ref{ffgeneric2}) leads to
\begin{equation}
\label{fff1}
\left(A-B\,\sqrt{1-\frac{R^2}{C^2}}\right)^2
=
1-\frac{2\,{\cal M}}{R}-\frac{\Lambda}{3}\,R^2
\ ,
\end{equation}
and
\begin{equation}
\label{fff2}
\left(1-\frac{R^2}{C^2}\right)
+\alpha\,f^*_R
=
1-\frac{2\,{\cal M}}{R}-\frac{\Lambda}{3}\,R^2
\ ,
\end{equation}
with $f^*_R=f^*(R)$ can be obtained from Eq.~\eqref{f}.
Continuity of the second fundamental form in Eq.~(\ref{matchingf}) under the constraint~\eqref{Constrain}
now holds provided 
\begin{equation}
\label{sfff}
p_R=0
\ ,
\end{equation}
which is ensured by the same expression for $A$ shown in Eq.~\eqref{Aconstant}.
This result is in agreement with our prescription to avoid having a solid crust and also ensures that
\begin{equation}
\label{f0}
f_R^{*}=0
\end{equation}
as we can verified from Eq.~\eqref{f}.
Therefore the condition~(\ref{fff2}) leads to 
\begin{equation}
\label{MASS}
{\cal M}=M
\ ,
\end{equation}
where Eq.~\eqref{Schwden} has been used.
The result in Eq.~\eqref{MASS} is in agreement with the expressions in~\eqref{ffgeneric2} and~\eqref{f0}
and tells us that the the constant $C$ is also related to $M$, $R$ and $\Lambda$ by the same expression
given in Eq.~\eqref{RCconstant}.
We remark that the condition~\eqref{f0} shows that, despite the anisotropic effects, the total mass of the
stellar distribution remains unchanged.
Finally, by using Eqs.~\eqref{Aconstant} and~(\ref{MASS}) in the matching condition~\eqref{fff1},
we obtain the same expression for the constant $B$ in Eq.~\eqref{Bconstant}.
We therefore conclude that the relations between the constants $A$, $B$ and $C$ in terms of $M$, $R$
and $\Lambda$ are not affected by the anisotropy.
\begin{figure}[t]
\center
\includegraphics[scale=0.8]{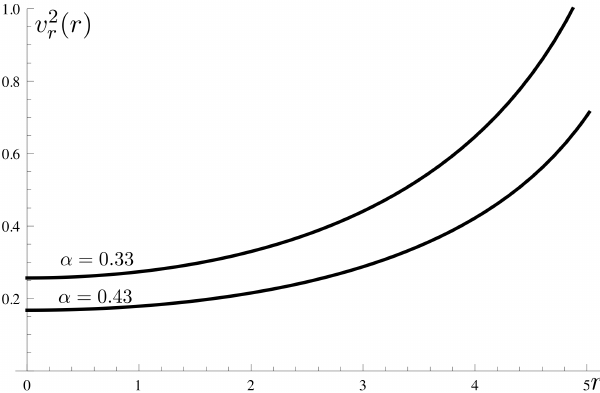}
\includegraphics[scale=0.8]{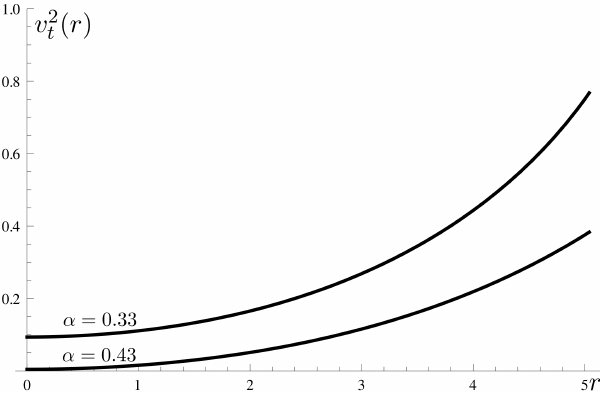}
\caption{Radial velocity ($v_r^2$) and tangential velocity ($v_t^2$) for a stellar system with compactness
$M/R=0.2$.
Radial coordinates are in units of $M$ ($r=5=R)$.
The cosmological constant $\Lambda\sim\,10^{-52}$ is too small to be relevant.}
\label{figv}      
\end{figure}
\par
Given a stellar distribution of mass $M$ and radius $R$ in a background with cosmological constant $\Lambda$,
we can now analyse the anisotropic effects on physical variables for different values of $\alpha$.
The first thing to notice is that the effective radial pressure $\tilde p_r$ remains proportional to the isotropic
expression~\eqref{Schwpre0}, and we therefore find the usual Buchdahl limit for the star
compactness~\cite{buchdahl} with cosmological constant in Eq.~\eqref{buch}.
This result is further supported by the fact that both the effective density $\tilde\rho$ and the tangential
pressure $\tilde p$ do not show any singularity for $0\le r\le R$ when Eq.~\eqref{buch} holds.
Moreover, the effective density is not uniform and $\tilde\rho'$ turns out to be proportional to $-\alpha$,
which implies that the effective density decreases (increases) towards the surface for $\alpha>0$ (respectively, 
$\alpha<0$).
In the following we shall therefore only consider cases with $\alpha>0$ so that the effective density decreases
from the centre outwards.
\par
Since the explicit expressions are rather cumbersome, Figure~\ref{figpr} shows the effective radial and tangential
pressures in Eqs.~(\ref{Schwpr}) and~(\ref{Schwpt}) for two values of $\alpha>0$ (compared to the standard
Schwarzschild solution $\alpha=0$).
We see that the anisotropy decreases the effective radial pressures $\tilde p_r$ for all values of $0\le r\le R$.
The effective tangential pressure $\tilde p_t$ is always smaller than $\tilde p_r$ in the centre of the star,
but decreases more slowly, and eventually becomes larger than $\tilde p_r$, near the surface.
Moreover, the tangential pressure at the surface $\tilde p_t(R)>0$, whereas $\tilde p_r(R)=p_R=0$ by
construction. 
On the other hand, we see from Figure~\ref{figrho} that the effective density $\tilde\rho$ in Eq.~\eqref{Schwd}
also shows a similar behaviour to $\tilde p_t$, and the anisotropy $\Pi$ increases steadily from the centre
(where $\Pi=0$) towards the surface.
We want to remark that both the dominant energy condition,
\begin{equation}
\tilde{\rho}
\geq
|\tilde{p}_r|
\ ,
\qquad
\tilde{\rho}
\geq
|\tilde{p}_t|\ ,
\end{equation}
and strong energy condition
\begin{equation}
\tilde{\rho}+\tilde{p}_r+2\,\tilde{p}_t
\geq
0
\ ,
\qquad
\tilde{\rho}+\tilde{p}_r
\geq
0
\ ,
\qquad
\tilde{\rho}+\tilde{p}_t
\geq
0
\end{equation}	
are satisfied~\cite{ponce}.
Finally, it is worth to examine the causal conditions,
\begin{eqnarray}
\label{causal}
&&
v_r^2(r)
=
\frac{d\tilde{p}_r(r,\alpha)}{d\tilde{\rho}(r,\alpha)}
\leq 1
\\
&&
v_t^2(r)
=
\frac{d\tilde{p}_t(r,\alpha)}{d\tilde{\rho}(r,\alpha)}
\leq 1
\ ,
\end{eqnarray}
where $v_r$ and $v_t$ are the radial and tangential sound speed, respectively,
displayed in Figure~\ref{figv}.
In both cases we observe that the speed decreases with the anisotropy and remains subluminal.
We also notice that both speeds increase from the center to the stellar surface, which might be
interpreted as a signal of instability.
This may be true in the case of perfect fluids, however, it is not necessarily so when anisotropic
effects are presents, as the stability conditions become much more involved~\cite{Luis1, Luis2}.
\section{Conclusions}
\label{con}
\setcounter{equation}{0}
By using the MGD-decoupling approach, we found the anisotropic and non-uniform version of
the Schwarzschild-de~Sitter interior solution with cosmological constant given by the
exact and analytical expressions displayed in Eqs.~(\ref{Schw00}), (\ref{schw11d}), and
(\ref{Schwd})-(\ref{Schwpt}).
Contrary to the well known Schwarzchild interior solution, this new system satisfies all of the
physical requirement, namely, it is regular at the origin, pressure and density are positive
everywhere (at least for $\alpha>0$), mass and radius are well defined when the Buchdahl
limit~\eqref{buch} holds, density and pressure decrease monotonically from the centre
outwards (for $\alpha>0$), the dominant energy condition is satisfied, and the sound speeds
are subluminal.
Regarding this last requirement, our interior anisotropic solution is causal, showing thus that
the anisotropic effects produce a more realistic stellar structure.
\par
The matching conditions at the stellar surface were studied in detail for an outer vacuum
Schwarzschild-de Siter space-time.
In particular, the continuity of the second fundamental form in Eq.~\eqref{sfgeneric} was shown
to yield the important result that the effective radial pressure $\tilde{p}_r$ can be made to vanish
at the surface by a suitable choice~\eqref{Constrain} of the anisotropic source.
The effective
pressure~\eqref{efecprera} contains both the isotropic pressure of the undeformed Schwarzschild
solution and the anisotropic effects produces by the inner geometric deformation $f^{*}$
induced by the generic energy-momentum $\theta_{\mu\nu}$, which could also represent
a specific matter source.
If the geometric deformation $f^{*}$ is positive and therefore weakens the gravitational field
[see Eq.~\eqref{effecmass}], an outer Schwarzschild-de Sitter vacuum could only be supported
if the isotropic $p_R < 0$ at the star surface, which can be interpreted as regular matter with
a solid crust~\cite{jo11}.
However, we could keep the standard condition $p_{R}=0$ by imposing that the anisostropic
effects on the radial effective pressure be proportional to $p(r)$, as shown in Eq.~\eqref{Constrain}.
This leads to a vanishing inner deformation $f^*_R=0$ and therefore the total mass $M$
of the standard Schwarzschild interior solution is not affected by the anisotropy, as we can see in the condition~\eqref{ffgeneric2}.
A direct consequence of this is that the surface redshift 
\be
z=\left[1-\frac{2\,{ M}}{R}\right]^{-1/2}-1
\ee
remain equal for both solutions, namely, the standard Schwarzschild solution and its new causal
version.
\par
In this paper we included a cosmological constant $\Lambda$ for generality.
However, since the present value of $\Lambda$ would introduce corrections
of order $10^{-52}$, its effects could be significant mainly at very large scales, with no sizeable
consequences on self-gravitating stellar objects.
In particular, it was shown~\cite{zdenek4} that $\Lambda$ plays a significant role for very
extended polytropic spheres that could describe galactic dark matter halos
(see also Ref.~\cite{zdenek5} where the effects of $\Lambda$ in several astrophysical situations
is summarised).
On the other hand, a large effective cosmological constant could be related to phase transitions
in the early universe, and that could influence compact objects created during this period,
like primordial black holes.
For example, the electroweak phase transition at $T_{\rm ew} \sim 100\,$GeV corresponds
to $\Lambda_{\rm ew} \sim 0.028\,$cm$^{-2}$,
while for the quark confinement at $T_{\rm qc} \sim 100\,$MeV one would have 
$\Lambda_{\rm qc} \sim 2.8\cdot 10^{-10}\,$cm$^{-2}$~\cite{zdenek6}.
\par
We conclude by mentioning that, for the study of non-primordial compact configurations,
we can safely ignore the cosmological constant without jeopardising our causal solution.
This yields even simpler expressions which could be exploited more easily to investigate
some interesting cases, such as the gravastar limit and the extended Kerr source~\cite{camilo1},
or even a possible generalisation including a nontrivial time dependence,
as those in the exact time-dependent version found in Ref.~\cite{gondolo},
but without the space-time singularities present therein.
\section{Acknowledgements}
J.O.~and S.Z.~have been supported by the Albert Einstein Centre for Gravitation and Astrophysics financed
by the Czech Science Agency Grant No.14-37086G and by the Silesian University at Opava internal
grant SGS/14/2016.
J.O.~thanks Luis Herrera for useful discussion and comments. 
L.G.~acknowledges the FPI Grant BES-2014-067939 from MINECO (Spain).
A.S.~is partially supported by Project Fondecyt 1161192, Chile and MINEDUC-UA project, code ANT 1855.
R.C.~is partially supported by the INFN grant FLAG and his work has also been carried out in the framework
of activities of the National Group of Mathematical Physics (GNFM, INdAM) and COST action {\em Cantata\/}. 

\end{document}